\documentclass[aps,pre,twocolumn,showpacs,superscriptaddress]{revtex4-2}

\usepackage{graphicx}
\usepackage{dcolumn}
\usepackage{bm}
\usepackage{amsmath,amssymb,mathtools}
\usepackage{color}
\usepackage{appendix}
\usepackage{xcolor}

\definecolor{darkblue}{rgb}{0,0,0.6}
\definecolor{darkred}{rgb}{0.6,0,0}
\definecolor{darkgreen}{rgb}{0,0.6,0}

\usepackage[colorlinks=true,urlcolor=darkblue,citecolor=darkblue,linkcolor=darkred,hyperfootnotes=false]{hyperref}

\usepackage{setspace}

\makeatother

\begin{document}

\title{$\alpha$-divergence improves the entropy production estimation via machine learning}

\author{Euijoon Kwon}
\affiliation{Department of Physics and Astronomy \& Center for Theoretical Physics, Seoul National University, Seoul 08826, Republic of Korea}

\author{Yongjoo Baek}
\email{y.baek@snu.ac.kr}
\affiliation{Department of Physics and Astronomy \& Center for Theoretical Physics, Seoul National University, Seoul 08826, Republic of Korea}

\date{\today}
 
\begin{abstract}
Recent years have seen a surge of interest in the algorithmic estimation of stochastic entropy production (EP) from trajectory data via machine learning. A crucial element of such algorithms is the identification of a loss function whose minimization guarantees the accurate EP estimation. In this study, we show that there exists a host of loss functions, namely those implementing a variational representation of the $\alpha$-divergence, which can be used for the EP estimation. By fixing $\alpha$ to a value between $-1$ and $0$, the $\alpha$-NEEP (Neural Estimator for Entropy Production) exhibits a much more robust performance against strong nonequilibrium driving or slow dynamics, which adversely affects the existing method based on the Kullback-Leibler divergence ($\alpha = 0$). In particular, the choice of $\alpha = -0.5$ tends to yield the optimal results. To corroborate our findings, we present an exactly solvable simplification of the EP estimation problem, whose loss function landscape and stochastic properties give deeper intuition into the robustness of the $\alpha$-NEEP.

\end{abstract}


\maketitle

\section{Introduction}
\label{sec:intro}

How irreversible does a process look? One may pose this question for two distinct reasons. First, whether a biological process requires energy dissipation is often a subject of much debate~\cite{Brangwynne2008,Weber2012}. To resolve this issue, it is useful to note that irreversibility suggests energy dissipation. Various hallmarks of irreversibility, such as the breaking of the fluctuation-dissipation theorem~\cite{Mizuno2007} and the presence of nonequilibrium probability currents in the phase space~\cite{Battle2016,Gladrow2016}, have been used to determine whether energy is dissipated. Second, whether a nonequilibrium system allows for an effective equilibrium description is an important issue. For instance, in active matter, despite the energy dissipation at the microscopic level, it has been argued that the large-scale phenomena allow for an effective equilibrium description~\cite{Tailleur2008,Speck2014,Takatori2014,Farage2015,Solon2018}. If we can quantify the irreversibility of an empirical process at various levels of coarse-graining~\cite{Fodor2016,Nardini2017}, it will provide us with helpful clues as to whether we should look for an effective equilibrium theory for the process.

 Based on the framework of stochastic thermodynamics, modern thermodynamics assigns entropy production (EP) to each stochastic trajectory based on its irreversibility~\cite{Seifert2012}. Thus, empirically measuring the irreversibility of a process is closely tied to the problem of estimating EP from sampled trajectories~\cite{Roldan2010,Roldan2012,Li2019,Vu2020,Otsubo2020,Otsubo2022,Lee2023,Kim2020}. A straightforward approach to the problem is to evaluate the relevant transition probabilities by directly counting the number of trajectory segments, which is called the {\em plug-in method}~\cite{Roldan2010,Roldan2012}. The method, readily applicable to discrete systems, can also be applied to continuous systems through the use of kernel functions~\cite{Li2019}. However, while this method is simple and intuitive, it requires a huge ensemble of lengthy trajectories for accurate estimations ({\em curse of dimensionality}). More recent studies proposed methods based on universal lower bounds of the average EP, such as the thermodynamic uncertainty relations~\cite{Li2019,Vu2020,Otsubo2020,Otsubo2022} and the entropic bound~\cite{Lee2023}. While these methods do not suffer from the curse of dimensionality and are applicable even to non-stationary processes~\cite{Otsubo2022,Lee2023}, their accuracy is impaired when the underlying bounds are not tight. Moreover, these methods are applicable only to the estimation of the average EP, not the EP of each trajectory.

Meanwhile, with the advent of machine learning techniques in physics, a novel method for EP estimation using artificial neural networks has been developed~\cite{Kim2020}. This method, called the Neural Estimator for Entropy Production (NEEP), minimizes the loss function based on a variational representation of the Kullback-Leibler (KL) divergence. Without any presupposed discretization of the phase space and using the rich expressivity of neural networks, the NEEP suffers far less from the complications of the sampling issues and is applicable to a diverse range of stochastic processes~\cite{Otsubo2022}.

Still, the NEEP has its limits. Its accuracy deteriorates when the nonequilibrium driving is strong or when the dynamics slows down so that the phase space is poorly sampled. In this study, we show that the NEEP can be significantly improved by changing the loss function. Toward this purpose, we propose the $\alpha$-NEEP, which generalizes the NEEP. Instead of the KL divergence, the $\alpha$-NEEP utilizes the $\alpha$-divergence, which has been mainly used in the machine learning community~\cite{Basu1998,Sugiyama2012,Nowozin2016,Belghazi2018}. We demonstrate that the $\alpha$-NEEP with nonzero values of $\alpha$ shows much more robust performance for a broader range of nonequilibrium driving and sampling quality, with $\alpha = -0.5$ showing the optimal performance overall. This is corroborated by an analytically tractable simplification of the $\alpha$-NEEP that shows the optimality of $\alpha = -0.5$.

The rest of this paper is organized as follows. After reviewing the original NEEP and its limitations (Sec.~\ref{sec:NEEP}), we introduce the $\alpha$-NEEP (Sec.~\ref{sec:alpha-NEEP}) and demonstrate its enhanced performance for three different examples of nonequilibrium systems (Sec.~\ref{sec:examples}). Then we investigate the rationale behind the observed results using a simplified model describing how the $\alpha$-NEEP works (Sec.~\ref{sec:gaussian_model}). Finally, we sum up the results and discuss their implications (Sec.~\ref{sec:summary}).

\begin{figure}
    \includegraphics[width=\columnwidth]{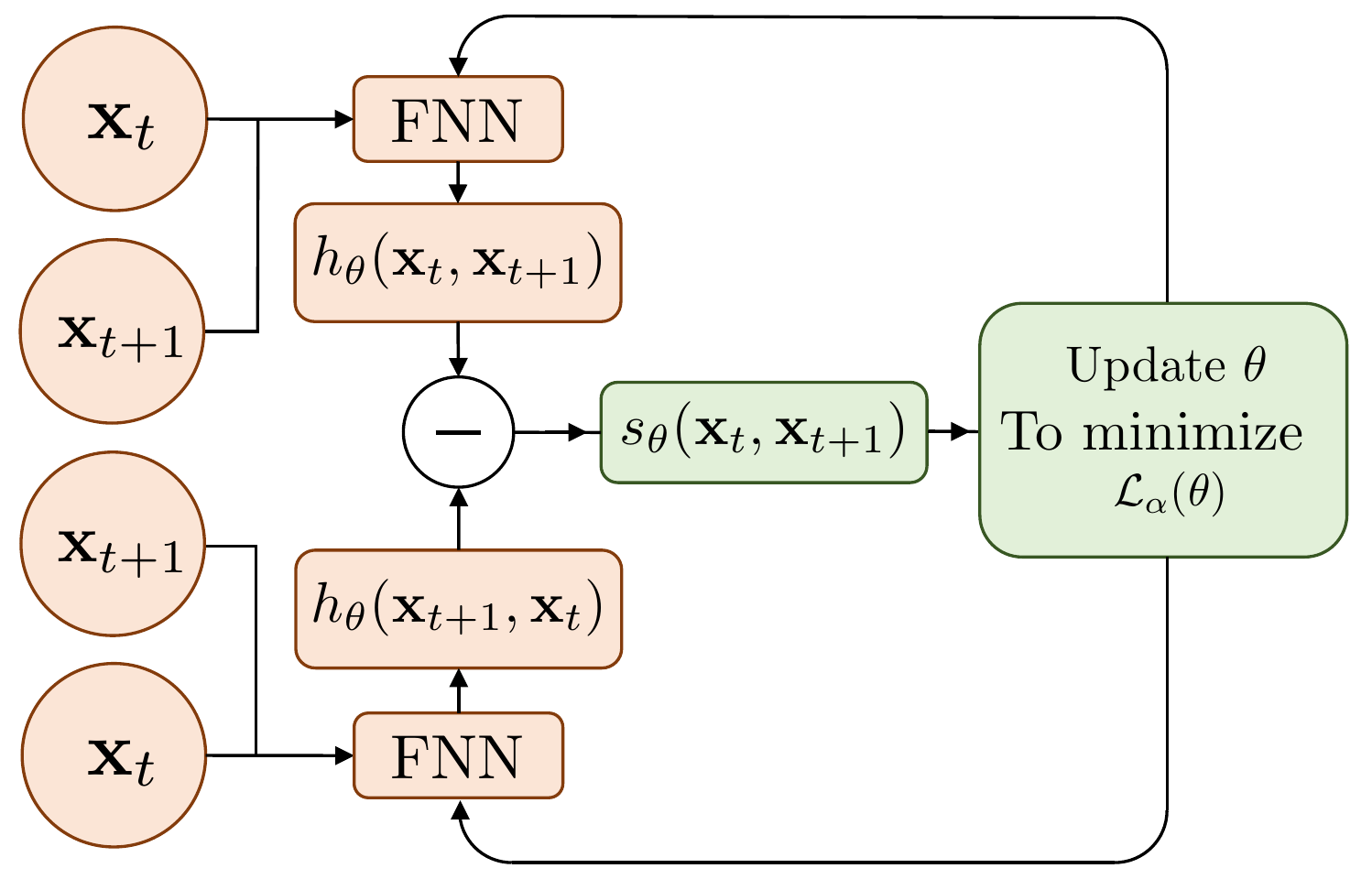}
    \caption{\label{fig:fig1} Schematic illustration of the neural-network implementation of the $\alpha$-NEEP.}
\end{figure}


\section{Overview of the Original NEEP}
\label{sec:NEEP}

We first give a brief overview of how the original NEEP~\cite{Kim2020} estimates EP at the trajectory level. Suppose our goal is to estimate EP of a Markov process in discretized time, $\mathbf{x}_t$, in a $d$-dimensional space. For every ordered pair of states, denoted by $\mathbf{x} \equiv (\mathbf{x}_t,\mathbf{x}_{t+1})$, there is EP associated with the transition between them, which is given by the ratio between the forward and the backward path probabilities
\begin{align} \label{eq:S_dev}
    \Delta S(\mathbf{x}) = \log \frac{p(\mathbf{x})}{p(\tilde{\mathbf{x}})},
\end{align}
where $\tilde{\mathbf{x}} \equiv (\mathbf{x}_{t+1}, \mathbf{x}_{t})$. Note that, throughout this study, we use the unit system in which the Boltzmann constant can be set to unity ($k_\mathrm{B} = 1$). Then it follows that the ensemble average of this EP is equivalent to the KL divergence, which satisfies the inequality
\begin{align} \label{eq:KL_div_var}
    \langle \Delta S(\mathbf{x}) \rangle &= D_{\rm{KL}}[p(\mathbf{x}),p(\tilde{\mathbf{x}})]
    \nonumber\\ & \ge \left \langle \log r(\mathbf{x}) - \frac{p(\tilde{\mathbf{x}})}{p(\mathbf{x})} r(\mathbf{x}) + 1 \right \rangle_{p(\mathbf{x})}
\end{align}
for any positive function $r(\mathbf{x})$, given that $\langle\cdot\rangle_{p(\mathbf{x})}$ denotes the average with respect to the distribution $p(\mathbf{x})$. This inequality can be proven as follows: since $\log$ is a concave function, the line tangent to any point never falls below the function. Thus, $\frac{1}{r_0} (r- r_0) + \log r_0 \ge \log r$ for any $r$ and $r_0$. By putting $r_0 = r_0(\mathbf{x}) = p(\mathbf{x})/p(\tilde{\mathbf{x}})$ and taking the average with respect to $p(\mathbf{x})$, we get the inequality. In this derivation, we immediately note that the equality condition is satisfied if and only if $r(\mathbf{x}) = r_0(\mathbf{x}) = p(\mathbf{x})/p(\tilde{\mathbf{x}})$. Hence, by varying $r(\mathbf{x})$ to maximize the right-hand side of Eq.~\eqref{eq:KL_div_var}, we accurately estimate the average EP $\langle\Delta S(\mathbf{x})\rangle$. For this reason, Eq.~\eqref{eq:KL_div_var} is called the {\em variational representation} of the KL divergence. Moreover, as a byproduct, we also obtain the function $r_0(\mathbf{x})$, which yields an accurate estimate for trajectory-level EP by $\Delta S(\mathbf{x}) = \log r_0(\mathbf{x})$.

Kim {\em et al.}~\cite{Kim2020} used these properties to construct the loss function of the NEEP. More specifically, they introduce $s_\theta (\mathbf{x})$, an estimator for trajectory-level EP parametrized by $\theta$, and put $r(\mathbf{x}) = e^{s_\theta (\mathbf{x})}$. Then, Eq.~\eqref{eq:KL_div_var} can be rewritten as
\begin{align}
    \langle \Delta S(\mathbf{x}) \rangle \ge \left \langle s_\theta (\mathbf{x}) - e^{s_\theta (\tilde{\mathbf{x}})} + 1 \right \rangle_{p(\mathbf{x})},
\end{align}
where $\langle r(\mathbf{x})\,p(\tilde{\mathbf{x}})/p(\mathbf{x}) \rangle_{p(\mathbf{x})} = \langle r(\mathbf{x}) \rangle_{p(\tilde{\mathbf{x}})} = \langle r(\tilde{\mathbf{x}}) \rangle_{p(\mathbf{x})}$ has been used based on the one-to-one correspondence between $\mathbf{x}$ and $\tilde{\mathbf{x}}$. Furthermore, since EP is odd under time reversal, {\em i.e.}, $\Delta S(\mathbf{x}) = -\Delta S(\tilde{\mathbf{x}})$, it is natural to impose the same condition on $s_\theta$. This leads to the inequality
\begin{align}
    \langle \Delta S(\mathbf{x}) \rangle \ge \left \langle s_\theta (\mathbf{x}) - e^{-s_\theta (\mathbf{x})} + 1 \right \rangle_{p(\mathbf{x})},
\end{align}
which motivates the loss function
\begin{align} \label{eq:NEEP}
    \mathcal{L}(\theta) = \left \langle - s_\theta (\mathbf{x}) + e^{-s_\theta (\mathbf{x})} - 1 \right \rangle_{p(\mathbf{x})},
\end{align}
so that the minimization of $\mathcal{L}(\theta)$ ensures the accurate EP estimation $s_\theta (\mathbf{x}) = \Delta S(\mathbf{x})$.

It is notable that $\mathcal{L}(\theta)$ defined above is a convex functional of $s_\theta$. Thus, as long as the $\theta$-dependence of $s_\theta$ is well behaved, any gradient-descent algorithm can reach the global minimum of $\mathcal{L}(\theta)$ without getting trapped in a local minimum. In this regard, the rugged loss function landscape is not a major issue of the NEEP.

However, the performance of the NEEP strongly depends on how well $p(\mathbf{x})$ is sampled. Since the second term of $\mathcal{L}(\theta)$ depends exponentially on $s_\theta(\mathbf{x})$, rare transitions with minute $p(\mathbf{x})$ can make nonnegligible contributions to $\mathcal{L}(\theta)$ when $e^{-s_\theta(\mathbf{x})}$ is extremely large. Since the frequency of rare events is subject to considerable sampling noise, the performance of the original NEEP deteriorates in the presence of a strong nonequilibrium driving which induces rare transitions with large negative EP. In the following section, we propose a loss function that remedies this weakness of the NEEP.


\section{Formulation of the $\alpha$-NEEP}
\label{sec:alpha-NEEP}

Here we formulate a generalization of the NEEP loss function with the goal of mitigating its strong sampling-noise dependence. We note that the loss function needs not be an estimator of average EP $\langle \Delta S(\mathbf{x}) \rangle$, for our goal is to estimate $\Delta S(\mathbf{x})$ at the level of each trajectory. Thus, while the original NEEP uses the variational representation of the KL divergence corresponding to $\langle \Delta S(\mathbf{x}) \rangle$, we propose a different approach based on the variational representation of the $\alpha$-divergence, which quantifies the difference between a pair of probability distributions $p(\mathbf{x})$ and $q(\mathbf{x})$ as
\begin{align}
D_\alpha[p\!:\!q] \equiv \left\langle\frac{[p(\mathbf{x})/q(\mathbf{x})]^{\alpha}-1}{\alpha(1+\alpha)}\right\rangle_{p(\mathbf{x})}.
\end{align}
Since this reduces to the KL divergence in the limit $\alpha \to 0$, our approach generalizes the NEEP by introducing an extra parameter $\alpha$. To emphasize this aspect, we term our method the $\alpha$-NEEP.

The goal of the $\alpha$-NEEP is to find $r(\mathbf{x})$ that minimizes the loss function
\begin{align} \label{eq:Lalpha_def}
    \mathcal{L}_\alpha[r] \equiv \left\langle-\frac{ r(\mathbf{x})^\alpha -1 }{\alpha} + \frac{q(\mathbf{x})}{p(\mathbf{x})} \frac{r(\mathbf{x})^{1+\alpha}-1}{1+\alpha} \right\rangle_{p(\mathbf{x})},
\end{align}
where $p(\mathbf{x})$ and $q(\mathbf{x})$ are probability density functions, 
and $\alpha$ is a real number other than $0$ and $-1$. See Appendix~\ref{app:DRE} for discussions of these two exceptional cases. It can be rigorously shown (see Appendix~\ref{app:DRE}) that $\mathcal{L}_\alpha[r]$ satisfies the inequality
\begin{align} \label{eq:Lalpha_div}
\mathcal{L}_\alpha[r] \ge -D_\alpha[p\!:\!q]
\end{align}
where the equality is achieved if and only if $r(\mathbf{x}) = p(\mathbf{x})/q(\mathbf{x})$ for all $\mathbf{x}$. In other words, by minimizing $\mathcal{L}_\alpha[r]$ to find $D_\alpha[p\!:\!q]$, we also obtain an estimate for the ratio $p(\mathbf{x})/q(\mathbf{x})$. We note that the properties of $\mathcal{L}_\alpha[r]$ used here are also valid for a much more general class of loss functions, as discussed in \cite{Basu1998,Sugiyama2012} (also see Appendix~\ref{app:DRE}).

Based on Eq.~\eqref{eq:Lalpha_div}, we can construct a loss function 
\begin{align} \label{eq:alphaNEEP_loss}
    \mathcal{L}_\alpha(\theta) = 
    \left\langle-\frac{e^{\alpha s_\theta (\mathbf{x})}-1}{\alpha}  + \frac{e^{-(1+\alpha)s_\theta (\mathbf{x})}-1}{1+\alpha}\right\rangle.
\end{align}
Note that this reduces to the loss function of the original NEEP shown in Eq.~\eqref{eq:NEEP} in the limit $\alpha \to 0$.
If $s_\theta$ is sufficiently well behaved, the minimization of $\mathcal{L}_\alpha(\theta)$ yields the minimizer $\theta^*$ which satisfies $\mathcal{L}_\alpha (\theta^*) = -D_\alpha [p(\mathbf{x}),p(\tilde{\mathbf{x}})]$ and $\Delta S(\mathbf{x}) = s_{\theta ^*} (\mathbf{x})$. The former is generally not equal to average EP $\langle \Delta S(\mathbf{x}) \rangle$ (unless $\alpha \to 0$), but the latter ensures the accurate estimation of trajectory-level EP $\Delta S(\mathbf{x})$.

Comparing Eqs.~\eqref{eq:NEEP} and \eqref{eq:alphaNEEP_loss}, one readily observes that the exponential dependence on $s_\theta(\mathbf{x})$ can be made much weaker in $\mathcal{L}_\alpha(\theta)$ by choosing the value of $\alpha$ between $-1$ and $0$. Since this mitigates the detrimental effects of the sampling error associated with rare trajectories with large negative $s_\theta(\mathbf{x})$, one can naturally expect that the performance of the $\alpha$-NEEP is much more robust against strong nonequilibrium driving. This is confirmed in the following sections.

Before proceeding, a few remarks are in order:
\begin{enumerate}
	\item The loss function satisfies $\mathcal{L}_{-(1+\alpha)} (\theta) = \mathcal{L}_\alpha (\theta)$, so the $\alpha$-NEEP is symmetric under the exchange $\alpha \leftrightarrow -(1+\alpha)$. For this reason, in the rest of this paper, we focus on the regime $-0.5 \le \alpha \le 0$ (the regime $\alpha > 0$ leads to very poor performance and is left out).
	\item From the antisymmetry $\Delta S(\mathbf{x}) = -\Delta S(\tilde{\mathbf{x}})$, we may set the estimator $s_\theta$ to be related to the feedforward neural network (FNN) output $h_\theta$ as
\begin{align}
	s_\theta(\mathbf{x}) = 	h_\theta(\mathbf{x})-h_\theta(\tilde{\mathbf{x}})\;,
\end{align}
so that the neural network focuses on the estimators that satisfy the antisymmetry of EP for more efficient training. The method described so far is schematically illustrated in Fig.~\ref{fig:fig1}.
\item We emphasize that the minimized $\mathcal{L}_\alpha$ is not directly related to average EP. In all cases, we compute the average EP by averaging $s_\theta$ over the sampled transitions.
\end{enumerate}


\section{Examples}
\label{sec:examples}

To assess the performance of the $\alpha$-NEEP for various values of $\alpha$, we apply the method to toy models of nonequilibrium systems, namely the two-bead model, the Brownian gyrator, and the driven Brownian particle.

\begin{figure}
    \includegraphics[width=\columnwidth]{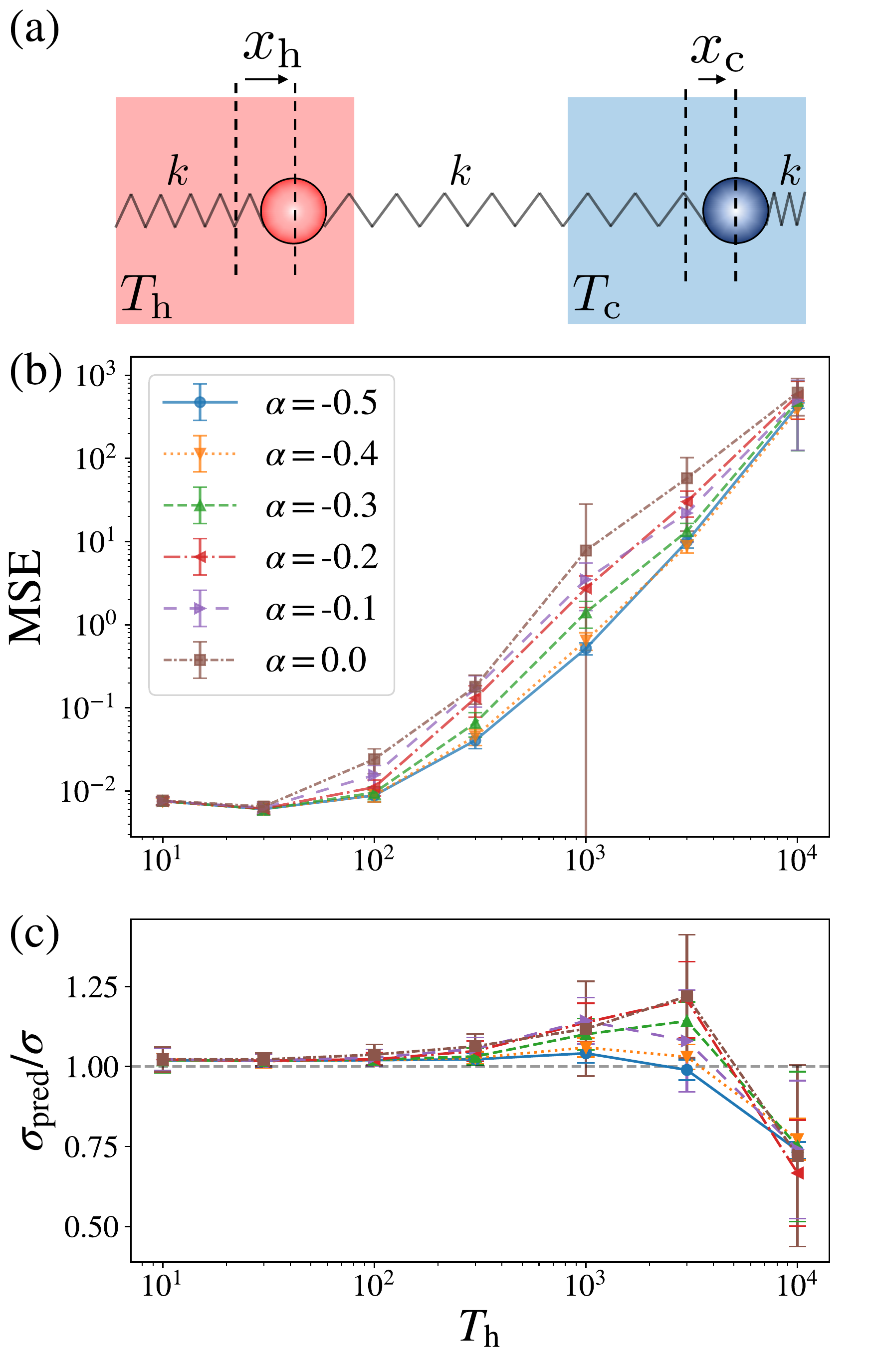}
    \caption{\label{fig:fig2}  (a) Illustration of the two-bead model. (b) Mean square error (MSE) of the EP estimate for various temperature differences. (c) Ratio between the estimated value $\sigma_\mathrm{pred}$ and the true value $\sigma$ of average EP for the two-bead model. Temperature of the cold bath is fixed at $T_\mathrm{c} = 1$. Each data point and error bar are obtained from $40$ independent trainings.}
    \end{figure}

(i) {\em The two-bead model}. This model has been used in a number of previous studies as a benchmark for testing EP estimators~\cite{Battle2016,Li2019,Otsubo2020,Kim2020}. The model consists of two one-dimensional (1D) overdamped beads which are connected to each other and to the walls on both sides by identical springs, see Fig.~\ref{fig:fig2}(a). The beads are in contact with heat baths at temperatures $T_{\rm{h}}$ and $T_{\rm{c}}$ with $T_{\rm{h}} > T_{\rm{c}}$. Denoting by $x_{\rm{h}}$ ($x_{\rm{c}}$) the bead in contact with the hot (cold) bath, the stochastic equations of motion are given by
\begin{subequations}\label{eq:two_bead_EOM}
	\begin{align}
    \gamma \dot {x}_{\rm{h}} &= k(-2x_{\rm{h}} + x_{\rm{c}}) + \sqrt{2\gamma T_{\rm{h}}} \xi_{\rm{h}}(t)\;, \\
    \gamma \dot {x}_{\rm{c}} &= k(-2x_{\rm{c}} + x_{\rm{h}}) + \sqrt{2\gamma T_{\rm{c}}} \xi_{\rm{c}}(t)
    \;.
\end{align}
\end{subequations}
Here $k$ is the spring constant, $\gamma$ the friction coefficient, and $\xi_{\rm{h},\,\rm{c}}$ the Gaussian thermal noise with zero means and $\langle\xi_{\rm{h}}(t)\xi_{\rm{h}}(t')\rangle=\langle\xi_{\rm{c}}(t)\xi_{\rm{c}}(t')\rangle=\delta(t-t')$. For infinitesimal displacements $(dx_{\rm{h}},dx_{\rm{c}})$, the associated EP is given by
\begin{align} \label{eq:two_bead_S}
    \Delta S = \frac{k}{\gamma}\left[\frac{2x_{\rm{h}} - x_{\rm{c}}}{T_{\rm{h}}} \circ dx_{\rm{h}} + \frac{2x_{\rm{c}} - x_{\rm{h}}}{T_{\rm{c}}} \circ dx_{\rm{c}} \right]  + \Delta S_{\text{sys}}
    \;,
\end{align}
where $\circ$ denotes the Stratonovich product and $\Delta S_\text{sys}$ the change of the system's Shannon entropy, namely
\begin{align}
\Delta S_\text{sys} = -\ln \frac{p_\mathrm{s}(x_{\rm{h}}+dx_{\rm{h}},x_{\rm{c}}+dx_{\rm{c}})}{p_\mathrm{s}(x_{\rm{h}},x_{\rm{c}})}
\end{align}
for the steady-state distribution $p_\mathrm{s}(x_{\rm{h}},x_{\rm{c}})$. Since the system is fully linear, $p_s(x_{\rm{h}},x_{\rm{c}})$ can be calculated analytically. Thus the EP of this model can be calculated exactly using Eq.~\eqref{eq:two_bead_S} and compared with the $\alpha$-NEEP result.

To see how the predicted EP differs from the true EP, we observe the behavior of the mean square error (MSE) $\langle(s_\theta - \Delta S)^2\rangle$. In Fig.~\ref{fig:fig2}(b), we observe that strengthening the nonequilibrium driving (by increasing $T_h$ while keeping $T_c = 1$) tends to impair the EP estimation. This is because a stronger driving makes the reverse trajectories of typical trajectories rarer, lowering the sample quality. The adverse effects of the nonequilibrium driving are the strongest for the original NEEP ($\alpha = 0$), which are mitigated by choosing different values of $\alpha$. Remarkably, choosing $\alpha = -0.5$ leads to the most robust performance against the driving.

As an alternative measure of the estimator's performance, we also observe the ratio between the predicted average EP $\sigma_\text{pred}$ and the exact average EP $\sigma$. The results are shown in Fig.~\ref{fig:fig2}(c), which exhibit two different regimes. As $T_h$ increases, there is a regime where the estimator overestimates average EP, which is followed by an underestimation regime. A detailed explanation for this behavior will be given in Sec.~\ref{sec:gaussian_model} using a simplified model. At the moment, we note that $\sigma_\text{pred}/\sigma$ tends to deviate away from $1$ most strongly for the original NEEP ($\alpha = 0$), while choosing different values of $\alpha$ makes the ratio stay closer to $1$. Again, the optimal value of $\alpha$ seems to be $-0.5$.

\begin{figure}
    \includegraphics[width=\columnwidth]{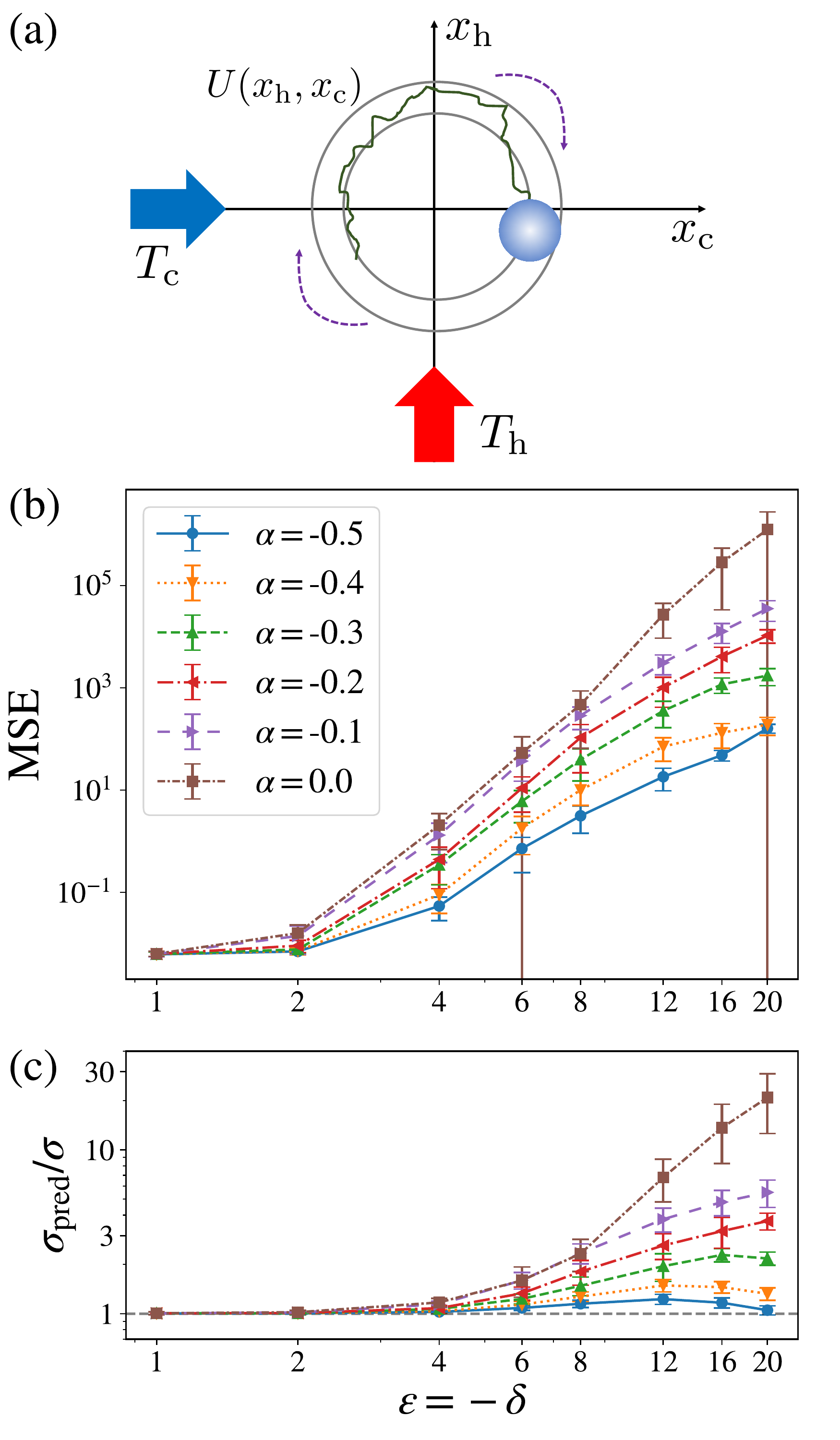}
    \caption{\label{fig:fig3}  (a) Illustration of the Brownian gyrator. Circles represent the equipotential lines and the dashed arrows indicate the directions of the nonconservative driving. (b) MSE of the EP estimate for the Brownian gyrator model as the magnitude of nonconservative force, $\varepsilon = -\delta$, is varied. (c) Ratio between the estimated value $\sigma_\mathrm{pred}$ and the true value $\sigma$ of average EP for the Brownian gyrator. Temperatures are fixed at $T_\mathrm{h} = 10$ and $T_\mathrm{c} = 1$. Each data point and error bar are obtained from $40$ independent trainings.}
    \end{figure}

(ii) {\em The Brownian gyrator.} This simple model of a single-particle heat engine allows us to check the effects of a nonequilibrium driving apart from the temperature difference $T_\mathrm{h} > T_\mathrm{c}$. The dynamics of the model is governed by
\begin{subequations}\label{eq:two_bead_EOM}
	\begin{align}
    \gamma \dot {x}_{\rm{h}} &= -\partial_{x_{\rm{h}}} U(x_{\rm{h}}, x_{\rm{c}}) + \varepsilon x_{\rm{c}} + \sqrt{2\gamma T_{\rm{h}}} \,\xi_{\rm{h}}(t)\;, \\
    \gamma \dot {x}_{\rm{c}} &= -\partial_{x_{\rm{c}}} U(x_{\rm{h}}, x_{\rm{c}}) + \delta x_{\rm{h}}+ \sqrt{2\gamma T_{\rm{c}}} \,\xi_{\rm{c}}(t)
    \;,
\end{align}
\end{subequations}
where $U(x_{\rm{h}}, x_{\rm{c}}) = \frac{1}{2} k (x_{\rm{h}} ^2 + x_{\rm{c}} ^2)$ is the harmonic potential, and $(\varepsilon x_\mathrm{c}, \delta x_\mathrm{h})$ is a nonconservative force that drives the system out of equilibrium and enables work extraction. See Fig.~\ref{fig:fig3}(a) for an illustration of this system. For infinitesimal displacements $(dx_{\rm{h}}, dx_{\rm{c}})$, the associated EP is given by
\begin{align}
    \Delta S = - \frac{Q_{\rm{h}}}{T_{\rm{h}}} - \frac{Q_{\rm{c}}}{T_{\rm{c}}} + \Delta S_{\text{sys}},     
\end{align}
where
\begin{subequations}
\begin{align}
    Q_{\rm{h}} &= (\partial_{x_{\rm{h}}} U - \epsilon x_c ) \circ dx_{\rm{h}},\\ Q_{\rm{c}} &= (\partial_{x_{\rm{c}}} U - \delta x_h ) \circ dx_{\rm{c}},
\end{align}
\end{subequations}
and $\Delta S_{\text{sys}}$ the change of the system entropy. Again, the system is fully linear and the steady-state distribution can be calculated analytically, allowing exact calculations of EP at the trajectory level.

Setting $T_{\mathrm{h}} / T_{\mathrm{c}} = 10$ and $\varepsilon = -\delta$, we vary the magnitude of $\varepsilon$ to assess the robustness of the $\alpha$-NEEP in terms of the MSE and the ratio $\sigma_\text{pred}/\sigma$, as shown in Figs.~\ref{fig:fig3}(b) and (c), respectively. The results are qualitatively similar to the case of the two-bead model: as the nonconservative driving gets stronger, the performance of the original NEEP ($\alpha = 0$) deteriorates the most, while other values of $\alpha$ yield more robust results. Again, $\alpha = -0.5$ seems to be the optimal choice.

\begin{figure}
    \includegraphics[width=\columnwidth]{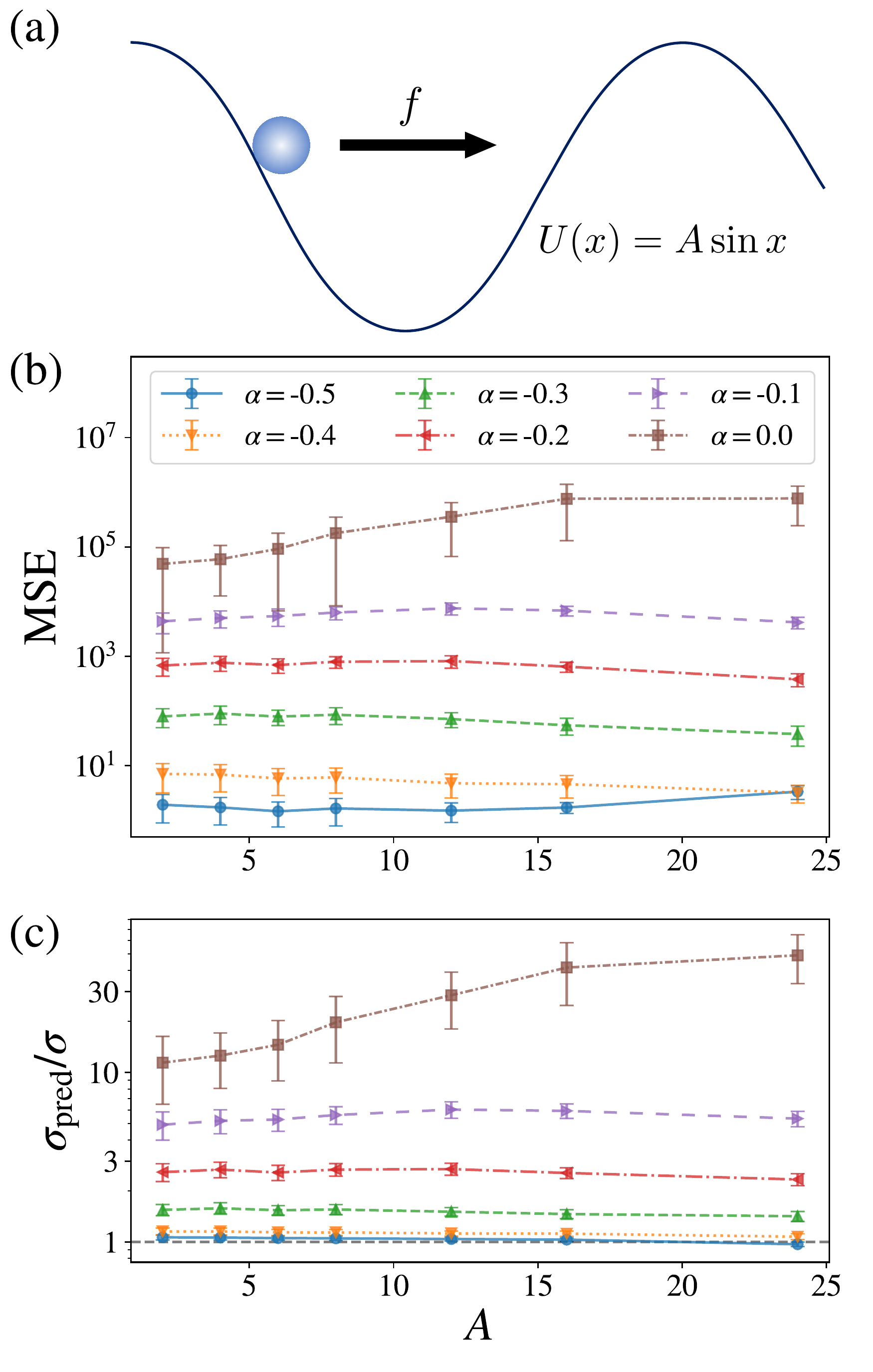}
    \caption{\label{fig:fig4}  (a) Illustration of the driven Brownian particle. (b) MSE of the EP estimate for the driven Brownian particle as the potential depth $A$ is varied. (c) Ratio between the estimated value $\sigma_\mathrm{pred}$ and the true value $\sigma$ of the average EP for the driven Brownian particle. Strength of the nonequilibrium driving is fixed at $f = 32$ and the temperature at $T = 1$. Each data point and error bar are obtained from $40$ independent trainings.}
    \end{figure}

(iii) {\em The driven Brownian particle}. While the two examples given above were both linear systems, we also consider a nonlinear system featuring a 1D overdamped Brownian particle in a periodic potential $U(x) = A \sin x$ driven by a constant force $f$. The motion of the particle is described by the Langevin equation
\begin{align} \label{eq:periodic}
    \gamma \dot x = f - U'(x) + \sqrt{2\gamma T} \xi(t)
    \;,
\end{align}
where $\xi(t)$ is a Gaussian white noise with unit variance. See Fig.~\ref{fig:fig4}(a) for an illustration of the model. For sufficiently large $A$, this model can approximate the behaviors of the Markov jump process on a discrete chain. For this model, the EP associated with the infinitesimal displacement $dx$ is given by
\begin{align}
\Delta S = \frac{-f\,dx + U(x+dx) - U(x)}{T} + \Delta S_\text{sys}\;,
\end{align}
where $\Delta S_\text{sys} = -\ln p_\mathrm{s}(x+dx)/p_\mathrm{s}(x)$ again denotes the Shannon entropy change for the steady-state distribution $p_\mathrm{s}(x)$. Since the system is 1D, it is straightforward to obtain $p_\mathrm{s}(x)$ by numerical integration. Thus, the EP of this model can also be calculated exactly and compared to the $\alpha$-NEEP result.

Fixing $f = 32$, the performance of the $\alpha$-NEEP for this model is shown in Figs.~\ref{fig:fig4}(b) and (c) in terms of the MSE and the ratio $\sigma_\mathrm{pred}/\sigma$, respectively. Due to the presence of a strong background driving ($f = 32$), there are already considerable differences among different methods at $A = 0$. But it is worth noting that increasing the amplitude $A$ of the periodic potential $U(x)$ clearly increases the MSE and makes $\sigma_\mathrm{pred}/\sigma$ deviate farther away from $1$ for the original NEEP ($\alpha = 0$). This may be the consequence of rarer movements (jumps from one potential well to the next) across the system as the potential well gets deeper, which means rare trajectories are even more poorly sampled. The $\alpha$-NEEPs with nonzero values of $\alpha$ are much more robust against the increase of $A$, with $\alpha = -0.5$ showing the best performance overall.

\begin{figure*}
    \includegraphics[width=\textwidth]{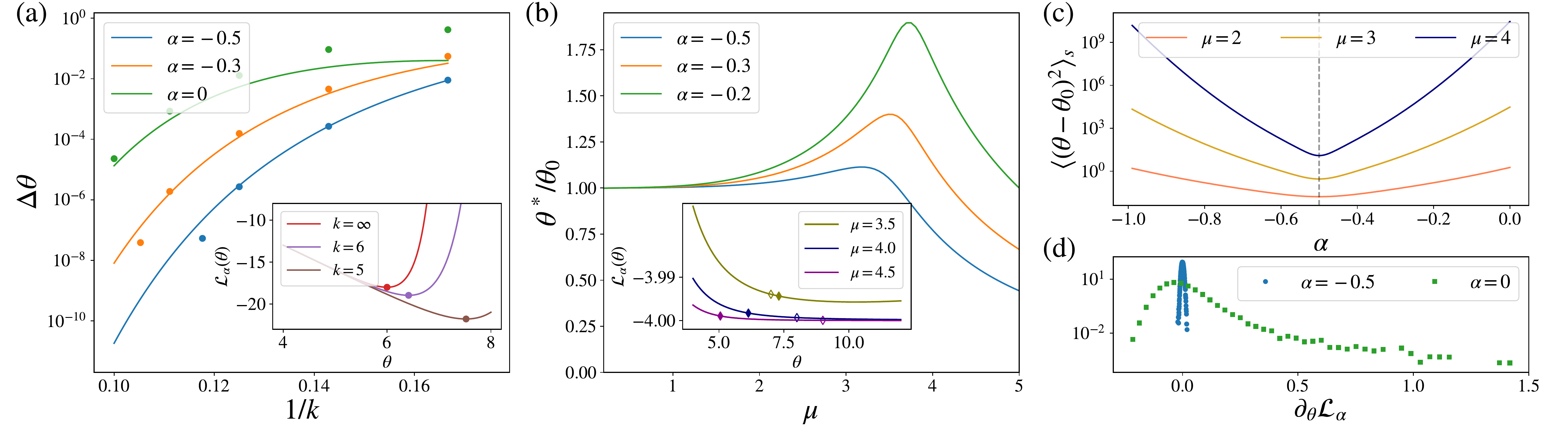}
    \caption{\label{fig:fig5} Performance of the exactly solvable one-parameter model. (a) Shift $\Delta \theta$ of the loss function minimum as a function of the truncation parameter $k$. Circles are results obtained by numerical minimization, and solid lines are from the small $1/k$ expansion. (Inset) Loss function landscapes, with circles indicating the minima. We fixed $\mu=3$, $\sigma=1$, and $\alpha=0$. (b) Ratio of the estimated minimum $\theta^*$ to the true minimum $\theta_0$ as the bias $\mu$ is varied. The optimal points are calculated using the criterion that the loss function gradient satisfies $|\partial_\theta \mathcal{L}_\alpha(\theta)| < 10^{-3}$ for the first time as $\theta$ increases from $0$. We fixed $k=4$ and $\sigma=1$. (Inset) Loss function landscape. Open diamonds indicate the true minima $\theta_0$, and the filled diamonds represent the estimated minima $\theta^*$. The parameters $\alpha = -0.5$ and $k=4$ are fixed. (c) MSE of $\theta$. The vertical dashed line shows that the error is minimized at $\alpha=-0.5$. (d) Distribution of the loss function gradient $\partial_\theta \mathcal{L}_\alpha$ at the minimum $\theta_0 = 2$ for $\mu = \sigma = 1$.}
    \end{figure*}



\section{Simple Gaussian Model} \label{sec:gaussian_model}

The results shown thus far clearly indicate that, by choosing a nonzero value of $\alpha$, the $\alpha$-NEEP can exhibit a much more robust performance against the adverse effects of the nonequilibrium driving. Moreover, $\alpha = -0.5$ seems to exhibit the best performance in many cases. To gain more intuition into these results, we simplify the EP estimation problem to the density-ratio estimation problem for a 1D random variable. To be specific, we estimate the log ratio $s(x)\equiv \log p(x)/p(-x)$ given samples drawn from the distribution $p(x)$. It is intuitively clear that this problem is structurally equivalent to EP estimation.



For further simplification, we set

\begin{align}
    p(x) = \begin{cases}
    \mathcal{N} \exp \left[- \frac{(x-\mu)^2}{2\sigma^2} \right] & \text{for $|x-\mu| \le k \sigma$,}
    \\ 0 & \text{otherwise.}
    \end{cases}
\end{align}
Here $\mathcal{N}$ is a suitable normalization factor, $\mu$ the positive mean of the distribution, $\sigma$ the width of the distribution, and $k$ a positive number truncating the tails of the distribution. While $k = \infty$ corresponds to the perfect sampling of a Gaussian distribution, a finite $k$ corresponds to the case where the tails of the distribution are poorly sampled.

For $k = \infty$, the correct answer to the problem is a linear function $s(x) = \theta_0 x$, where $\theta_0 \equiv 2\mu/\sigma^2$. Thus, for further simplicity, we focus on the {\em one-parameter model} $s_\theta(x) = \theta x$, which estimates $s(x)$ using only a single parameter $\theta$. For this problem, the suitable loss function is obtained as an analog of Eq.~\eqref{eq:alphaNEEP_loss}:
\begin{align} \label{eq:loss_one_parameter}
    \mathcal{L}_\alpha(\theta) = \left\langle-\frac{(1+\alpha) e^{\alpha \theta x}-1}{\alpha} + e^{-(1+\alpha)\theta x} \right\rangle_{p(x)}
    \;.
\end{align}

If $k$ is large but finite, the minimum of this loss function shifts to $\theta_0 + \Delta\theta$, where $\Delta\theta$ can be expanded to the leading orders in $1/k$:
\begin{align}
    \Delta \theta \sim \exp \left[- \frac{k^2}{2} + \frac{2\mu}{\sigma} \left(\left| \alpha + \frac{1}{2}\right| + \frac{1}{2} \right)k \right]
    \;.
\end{align}
This clearly shows that $\alpha = -0.5$ gives the least shift $\Delta\theta$, as also illustrated by various results shown in Fig.~\ref{fig:fig5}.

In Fig.~\ref{fig:fig5}(a), we show that the shift of the minimum $\Delta\theta$ tends to increase as the tail sampling becomes poorer ({\em i.e.}, $k$ decreases). The landscapes of the loss function $\mathcal{L}_\alpha(\theta)$, shown in the inset of Fig.~\ref{fig:fig5}(a), also confirm this observation. The increase of the error with the potential depth $A$ in Figs.~\ref{fig:fig1}(d) and \ref{fig:fig2}(b) may primarily be due to the same effect. 

In Fig.~\ref{fig:fig5}(b), we plot the ratio between the estimated minimum $\theta^*$ and the true minimum $\theta_0$ as a function of the mean $\mu$, which is an analog of the nonequilibrium driving. We note that here $\theta^*$ is the lowest value of $\theta$ at which the slope of the loss function becomes less then $10^{-3}$. We observe that an overestimation regime ($\theta^*/\theta_0 > 1$) crosses over to an underestimation regime ($\theta^*/\theta_0 < 1$) as $\mu$ grows. This is in striking agreement with the trends shown in Fig.~\ref{fig:fig2}(a). The reason why $\theta^*$ underestimates $\theta_0$ for large $\mu$ can be understood by the flattened loss function landscapes shown in the inset of Fig.~\ref{fig:fig5}(b). In this regime, the dynamics of $\theta$ (starting from $\theta$ = 0) slows down, ending up at a value (filled diamonds) even lower than $\theta_0$ (empty diamonds). This effect is due to the samples with $x < 0$ vanishing when $\mu$ is too large. We expect that a similar mechanism might be at play behind the observed behavior of $\sigma_\mathrm{pred}/\sigma$ shown in Fig.~\ref{fig:fig2}(a). If we had used a broader range of nonequilibrium driving, the same behaviors might have been observed for other models as well, although this remains to be checked.

The one-parameter model also allows us to examine the effects of the finite minibatch size $M$. While the ideal loss function is given in Eq.~\eqref{eq:loss_one_parameter}, the loss function used in the actual training looks like
\begin{align}
    \mathcal{L}_\alpha(\theta; M) = \frac{1}{M} \sum_i{ \left[-\frac{(1+\alpha)e^{\alpha \theta X_i}-1}{\alpha} + e^{-(1+\alpha)\theta X_i}\right]}
    \;,
\end{align}
where $X_1,\,\ldots,\,X_M$ are i.i.d. Gaussian random variables of mean $\mu$ and variance $\sigma^2$. When $M$ is large and finite, using the central limit theorem (CLT), the gradient of this loss function can be approximated as~\cite{Li2017,Chaudhari2018}
\begin{align}
    \frac{\partial \mathcal{L}_\alpha}{\partial \theta}\bigg| _{\theta = \theta_t} = \bar{K} (\theta_t - \theta_0) + \sqrt{\frac{\Lambda}{M}} N_t + o(M^{-1/2})
    \;,
\end{align}
where $\theta_0 = \text{argmin}(\langle\mathcal{L}_\alpha (\theta) \rangle)$, $\bar{K} = \partial^2 _\theta \langle\mathcal{L}_\alpha\rangle \big|_{\theta = \theta_0}$, and $\Lambda = \text{Var}\big[\partial_\theta\mathcal{L}_\alpha \big|_{\theta=\theta_0,\,M=1} \big]$. When the stochastic gradient descent $\theta_{t+1} = \theta_t - \lambda (\partial\mathcal{L}_\alpha/\partial\theta) \big|_{\theta=\theta_t}$ reaches the steady state, the MSE of $\theta$ is given by
\begin{align}
    \langle \left( \theta - \theta_0 \right)^2 \rangle_s = \frac{\lambda \Lambda}{M \bar{K} (2-\lambda \bar{K})} + o(M^{-1})
    \;.
\end{align}
This leading-order behavior is shown in Fig.~\ref{fig:fig5}(c) for various values of $\mu$. For all cases, the MSE of $\theta$ is minimized at $\alpha = -0.5$, which is consistent with the smallest error bars observed at $\alpha = -0.5$ in Figs.~\ref{fig:fig1} and \ref{fig:fig2}. Hence, $\alpha = -0.5$ yields the most consistent EP estimator.

Direct measurements of the loss function gradient at the minimum also confirm the above result. As shown in Fig.~\ref{fig:fig5}(d), the gradient $\partial_\theta \mathcal{L}_\alpha$ is far more broadly distributed for $\alpha = 0$ than for $\alpha = -0.5$. Moreover, due to the subleading effects (beyond the CLT) of finite $M$, the gradient for $\alpha = 0$ features a large skewness. These show that the training dynamics for the original NEEP ($\alpha = 0$) tends to be far more volatile and unstable than for the $\alpha$-NEEP with $\alpha = -0.5$.



\section{Summary and outlook}
\label{sec:summary}

We proposed the $\alpha$-NEEP, a generalization of the NEEP for estimating steady-state EP at the trajectory level. By choosing a value of $\alpha$ between $-1$ and $0$, the $\alpha$-NEEP weakens the exponential dependence of the loss function on the EP estimator, effectively mitigating the adverse effects induced by poor sampling of transitions associated with large negative EP in the presence of strong nonequilibrium driving and/or deep potential wells. We also observed that $\alpha = -0.5$ tends to exhibit the optimal performance, which can be understood via a simplification of the original EP estimation problem, whose loss function landscape and relaxation properties are analytically tractable. The $\alpha$-NEEP thus provides a powerful method for estimating the EP for much broader range of the nonequilibrium driving force and the time scale of dynamics. Identification of even better loss functions and optimization of other hyperparameters (network size, number of iterations, etc.) are left as future works. It would also be interesting to apply the $\alpha$-NEEP to estimations of the EP of the Brownian movies~\cite{Bae2021} and stochastic systems with odd-parity variables~\cite{Kim2021}, which have been studied using the original NEEP method.

{\em Acknowledgments.} --- This work was supported by the POSCO Science Fellowship of the POSCO TJ Park Foundation. E.K. and Y.B. also thank Junghyo Jo and Sangyun Lee for helpful comments.

\begin{appendix}


\begin{figure}
\includegraphics[width=0.99\columnwidth]{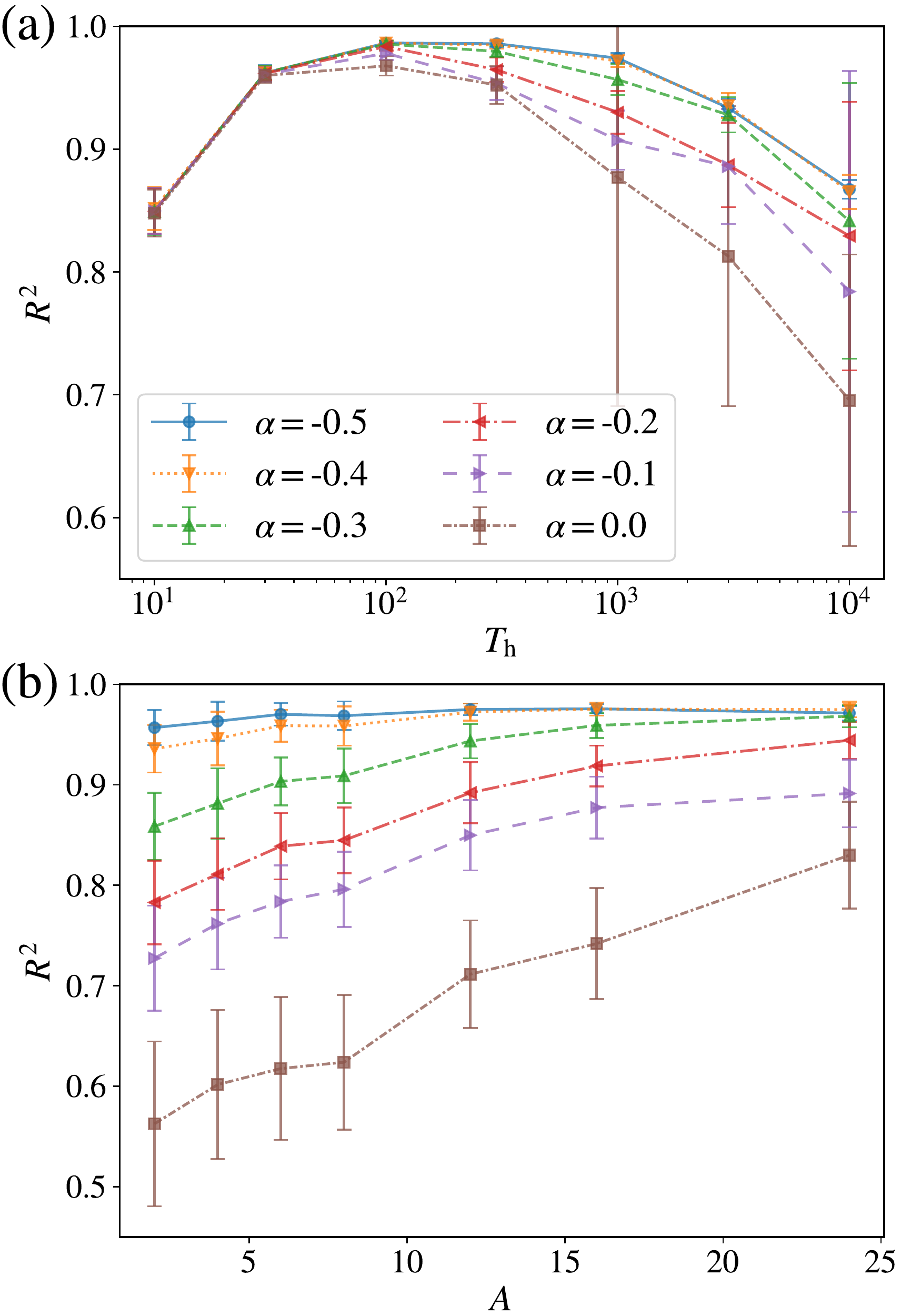}
\caption{\label{fig:figS1} Coefficient of determination $R^2$ for (a) the two-bead model with $T_\mathrm{c} = 1$ and for (b) the driven Brownian particle with $f = 32$. $10^4$ trajectories are used for each minibatch, and error bars indicate the standard deviations obtained from $40$ independent trainings.}
\end{figure}

\begin{figure}
\includegraphics[width=0.99\columnwidth]{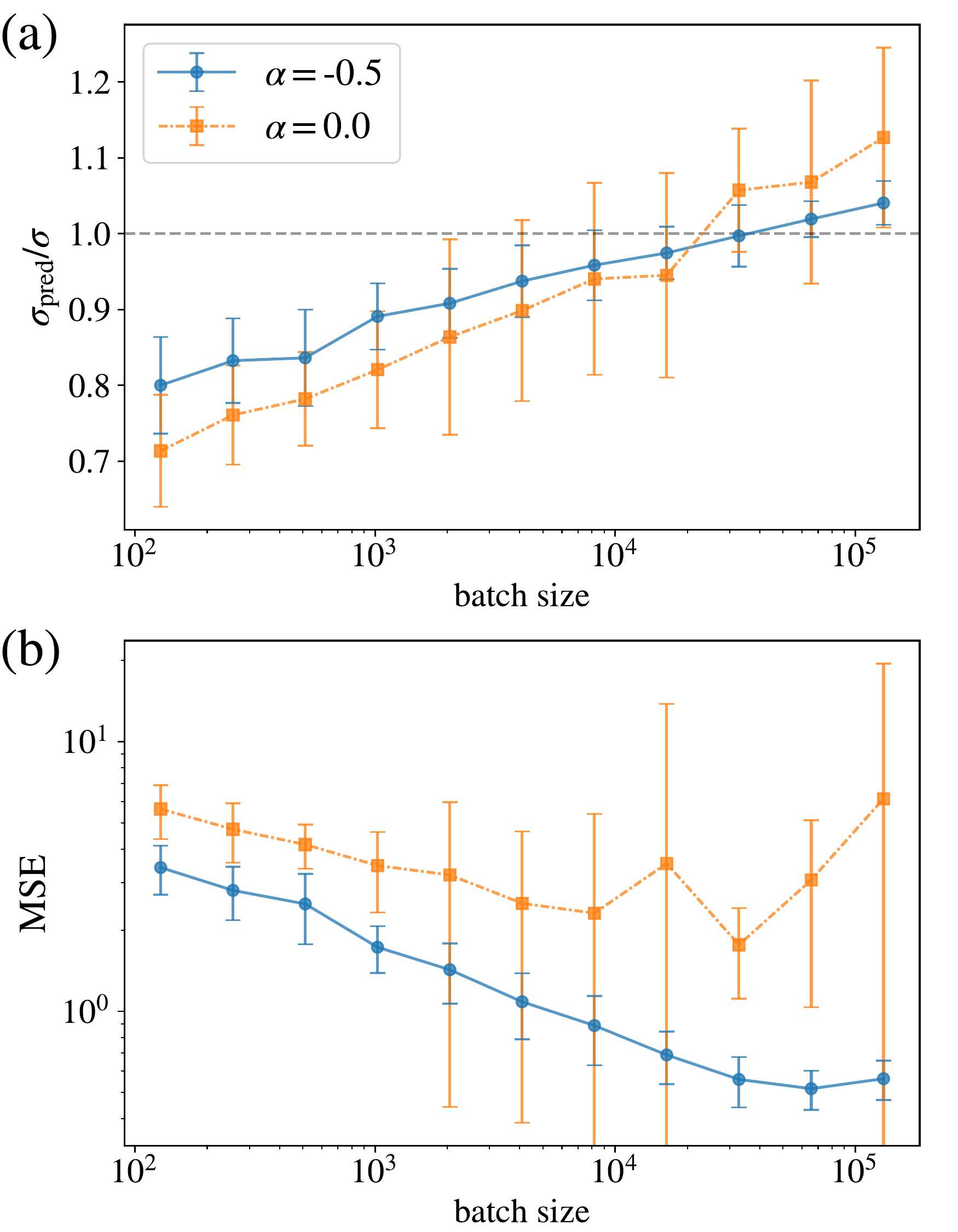}
    \caption{\label{fig:figS2} Effects of the minibatch size on the performance of the $\alpha$-NEEP for the two-bead model with $T_\mathrm{h} =1000$ and $T_\mathrm{c} = 1$. Error bars indicate the standard deviations obtained from $40$ independent trainings.}
\end{figure}

\section{Training details}
\label{app:training}

We always use the fully connected network (FCN) with three hidden layers, with each layer composed of $512$ nodes. Each training dataset consists of $10^6$ trajectories. The neural network parameters are updated using the ReLU activation function and the Adam optimizer. The learning rate is fixed to $10^{-5}$ and the weight decay is fixed to $5\times 10^{-5}$. We halt the training after $10000$ iterations, except for the results shown in Figs.~\ref{fig:figS3} and \ref{fig:figS4} (see Appendix~\ref{app:extra}), where we continue the training for a longer time to check the overfitting effects. All trainings are done on PyTorch with NVIDIA GeForce RTX 3090.
    
In subfigure (b) of Figs.~\ref{fig:fig2}--\ref{fig:fig4}, each minibatch consists of $10^4$ trajectories. On the other hand, in subfigure (c) of Figs.~\ref{fig:fig2}--\ref{fig:fig4}, each minibatch consists of $10^5$ trajectories.

\section{Density ratio estimation via $f$-divergence}
\label{app:DRE}

Here we show that the the loss function $\mathcal{L}_\alpha[r]$ given in Eq.~\eqref{eq:Lalpha_def}, whose minimization allows us to estimate the ratio between two probability density functions, can be generalized even further using the concept of $f$-divergence. Consider a convex, twice-differentiable real-valued function $f(u)$. Then, the inequality 
\begin{align} \label{eq:fineq}
-pf'(u)+q \left[uf'(u) - f(u) \right] \ge -qf(p/q)
\;
\end{align}
holds. We can verify this by differentiating the left-hand side (LHS) with respect to $u$, which yields $f''(u)(-p+qu)$. Thus, the LHS has a local minimum at $u=p/q$, and this is the only local minimum since $f$ is convex. In addition, the second derivative of the LHS at $u=p/q$ equals $qf''(p/q)$, which is positive by the convexity. This proves the inequality~\eqref{eq:fineq}.

Using this result, we can design a loss function whose minimum is equal to the negative $f$-divergence between two probability distributions $p(\mathbf{x})$ and $q(\mathbf{x})$. To be specific, for any function $r(\mathbf{x})$, we define
\begin{align} \label{eq:fdiv_rep}
    \mathcal{L}_f[r] &= \left\langle -f'(r(\mathbf{x})) + \frac{q(\mathbf{x})}{p(\mathbf{x})} \left[r(\mathbf{x}) f'(r(\mathbf{x})) - f(r(\mathbf{x})) \right] \right\rangle _{p(\mathbf{x})} \nonumber
    \\ &=\int d^{2d} \mathbf{x} \,\{-p(\mathbf{x}) f'\!\left(r (\mathbf{x})\right)\nonumber\\
    &\quad \quad\quad\quad\quad+ q(\mathbf{x}) \left[r(\mathbf{x}) f'\!\left(r(\mathbf{x})\right)-f\!\left(r (\mathbf{x})\right) \right] \} \;.
\end{align}

Using Eq.~\eqref{eq:fdiv_rep}, we conclude that 
\begin{align} \label{eq:fdiv_rep_pf}
    \mathcal{L}_f [r] \ge -\int d^{2d} \mathbf{x}\, q(\mathbf{x})\, f\!\left( \frac{p(\mathbf{x})}{q(\mathbf{x})} \right) = -D_f[p\!:\!q]
    \;,
\end{align}
where $D_f[p{{\,:\,}}  q]$ is the $f$-divergence between the distributions $p(\mathbf{x})$ and $q(\mathbf{x})$, and the equality holds if and only if $r(\mathbf{x}) = p(\mathbf{x})/q(\mathbf{x})$ for all $\mathbf{x}$. By minimizing $\mathcal{L}_f[r]$, we can estimate $p(\mathbf{x})/q(\mathbf{x})$ as well as $D_f[p\!:\!q]$.

The loss function $\mathcal{L}_\alpha$ and the associated $\alpha$-divergence discussed in the main text are obtained by choosing the function $f$ to be
\begin{align} \label{eq:f_alpha}
    f_\alpha (u) = 
    \begin{cases}
    \frac{u^{1+\alpha}-(1+\alpha)u+\alpha}{\alpha (1+\alpha)} & \text{for $\alpha \neq 0,\,-1$,} \\
    u\log u & \text{for $\alpha = 0$,} \\
    \log u + 1 - u & \text{for $\alpha = -1$.}
    \end{cases}
    \;.
\end{align}
Note that $f_0(u) = \lim_{\alpha \to 0} f_\alpha(u)$ and $f_{-1}(u) = \lim_{\alpha \to -1} f_\alpha(u)$. It is straightforward to obtain Eq.~\eqref{eq:alphaNEEP_loss} and its extensions to the cases $\alpha = 0$ and $\alpha = -1$ from this choice.

\begin{figure*}
    \includegraphics[width=0.99\textwidth]{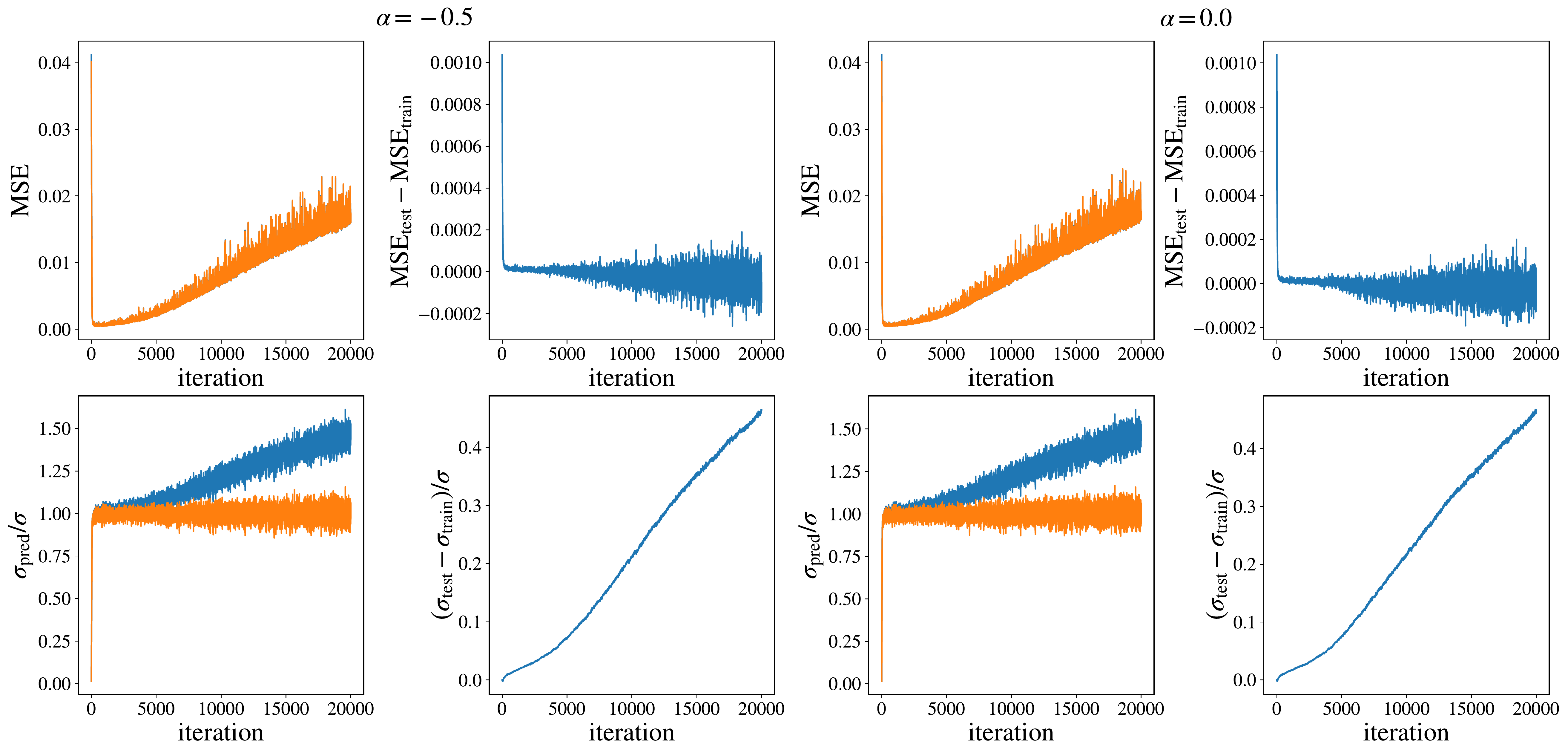}
    \caption{\label{fig:figS3} Training dynamics of the $\alpha$-NEEP for the two-bead model at $T_\mathrm{h} = 10$. Each minibatch consists of $10^5$ trajectories. The first and the third columns show the performance for the training set (blue/dark gray curve) and the test set (orange/light gray curve). The second and the fourth columns show the difference in performance between the two datasets. The first (last) two columns correspond to $\alpha = -0.5$ ($\alpha = 0$).} 
\end{figure*}

\begin{figure*}	\includegraphics[width=0.99\textwidth]{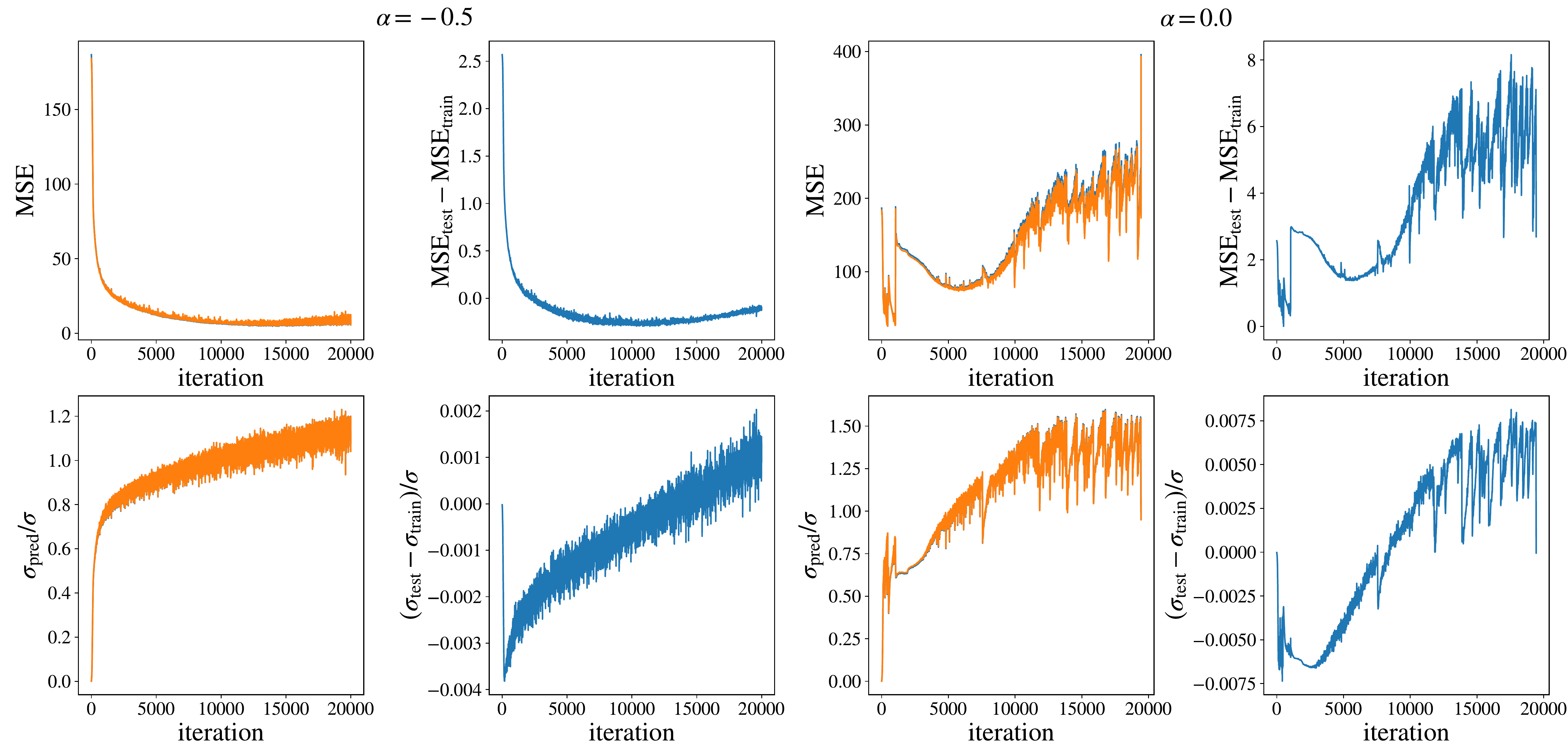}
    \caption{\label{fig:figS4}
    Training dynamics of the $\alpha$-NEEP for the two-bead model at $T_\mathrm{h} = 3000$ and $T_\mathrm{c} = 1$. Each minibatch consists of $10^5$ trajectories. The first and the third columns show the performance for the training set (blue/dark gray curve) and the test set (orange/light gray curve). The second and the fourth columns show the difference in performance between the two datasets. The first (last) two columns correspond to $\alpha = -0.5$ ($\alpha = 0$).}
\end{figure*}

\section{Extra numerical results} \label{app:extra}

\subsection{Coefficient of determination}

In the literature, the extent of agreement between a prediction and the true value is often expressed by the coefficient of determination $R^2$. Here we check how the behaviors of $R^2$ differ as the value of $\alpha$ changes for the cases of the two-bead model and the driven 1D Brownian particle.

 For the two-bead model, as shown in Fig.~\ref{fig:figS1}(a), $R^2$ exhibits a nonmonotonic behavior as a function of $T_\mathrm{h}$. The decrease of $R^2$ with increasing $T_\mathrm{h}$ reflects the detriment of the $\alpha$-NEEP performance as the nonequilibrium driving gets stronger. Meanwhile, the decrease of $R^2$ as $T_\mathrm{h}$ decreases (getting closer to equilibrium $T_\mathrm{h} = T_\mathrm{c} = 1$) is due to the overfitting phenomenon discussed in the next section, which disrupts the linear relationship between the predicted EP and the true EP.
 
 For the driven Brownian particle, as shown in Fig.~\ref{fig:figS1}(b), $R^2$ always increases with $A$. This may seem contradictory to how the MSE tends to increase or stay constant with increasing $A$ in Fig.~\ref{fig:fig4}(b). Indeed, higher $R^2$ only means that there is a good linear relationship between the EP estimate $s_\theta$ and the true EP $\Delta S$, not that $s_\theta$ and $\Delta S$ are close to each other. When $A$ is increased, due to the slower dynamics, we may have $s_\theta > \Delta S$ for transitions with positive EP and $s_\theta < \Delta S$ for transitions with negative EP, which can make the linear relationship between $s_\theta$ and $\Delta S$ appear stronger. This example clearly shows that $R^2$ is not an adequate measure of the performance of EP estimators.

\subsection{Effects of the minibatch size}

The minibatch refers to the group of samples used for computing the gradient of the loss function. Smaller (larger) minibatches increase (decrease) the noisy component of the gradient, which in turn affects the performance of the $\alpha$-NEEP.

We explicitly check the effects of the minibatch size using the two-bead model with $T_\mathrm{h} = 1000$ and $T_\mathrm{c} = 1$, as shown in Fig.~\ref{fig:figS2}. We use the ratio $\sigma_\mathrm{pred}/\sigma$ and the MSE as two different measures of the $\alpha$-NEEP performance. For small minibatches, the highly skewed distribution of the stochastic gradient shown in Fig.~\ref{fig:fig5}(d) causes underestimation of the EP. For large minibatches, the noisy component of the loss-function gradient decreases, revealing the properties of the loss function landscape of the training dataset. As discussed using the Gaussian model in Sec.~\ref{sec:gaussian_model}, the loss function landscape at a moderately strong nonequilibrium driving leads to the overestimation of the EP. Thus, as the minibatch size is increased, $\sigma_\mathrm{pred}/\sigma$ grows beyond $1$.

The nonmonotonic behaviors of the MSE also hint at the existence of an optimal minibatch size at the tradeoff between the skewed noise in the gradient (which drives the neural network towards underestimation) and the loss function landscape tilted towards overestimation. For both measures, the superiority of $\alpha = -0.5$ to $\alpha = 0$ is manifest.

\subsection{Effects of overfitting}

In many cases, when the training continues for too many iterations, artificial neural networks are known to exhibit overfitting behaviors. As shown in Figs.~\ref{fig:figS3} and \ref{fig:figS4}, we checked whether the $\alpha$-NEEP is also subject to the same phenomena as the training continues up to $20000$ iterations. Towards this end, we created two independent datasets of trajectories exhibited by the two-bead model, namely the training set and the test set. Only the former was used during the training of the $\alpha$-NEEP, and we measured the MSE and the ratio $\sigma_\mathrm{pred}/\sigma$ to assess the performance of the $\alpha$-NEEP for each dataset.

 In Fig.~\ref{fig:figS3}, we show the results for the weak nonequilibrium driving ($T_\mathrm{h} = 10$ and $T_\mathrm{c} = 1$). The first and the third columns show the two different measures of performance for the training dataset and the test dataset. Meanwhile, the second and the fourth columns show the difference between the corresponding measures obtained for two datasets. The overfitting phenomena are manifest from the increase of the MSE towards the end of the training. Interestingly, overfitting leads to an overestimation of the average EP only for the training dataset. We also note that the value of $\alpha$ is largely irrelevant to the extent of overfitting. This phenomenon can be explained as follows. Near equilibrium, the neural network swiftly reaches the loss function minimum. However, as the training continues, the neural network starts to see the detailed fluctuations of the training dataset. This makes the functional form of the estimator $s_\theta$ very rough, leading to the increase of the MSE for both datasets. But while the neural network now believes all trajectories in the training dataset to be highly irreversible and assigns high EP to them, the EP assigned to the trajectories in the test dataset stay unbiased. Thus, $\sigma_\mathrm{pred}/\sigma$ grows larger only for the training dataset.
 
 In Fig.~\ref{fig:figS4}, we show the results for the strong nonequilibrium driving ($T_\mathrm{h} = 3000$ and $T_\mathrm{c} = 1$). The subfigures are organized in exactly the same way as in Fig.~\ref{fig:figS3}. In this case, the overfitting effects do exist. But they are not as pronounced as in the case of the weaker nonequilibrium driving, and the differences between the training and the test datasets stay small. Note that the curves for $\alpha = 0$ exhibit strong fluctuations, which is in agreement with the large fluctuations of the gradient shown in Fig.~\ref{fig:fig5}(d).


\end{appendix}

\bibliography{alphaNEEP}

\end{document}